\documentclass[conference, letterpaper]{IEEEtran}
\IEEEoverridecommandlockouts
\usepackage{cite}
\usepackage{amsmath,amssymb,amsfonts}
\usepackage{algorithmic}
\usepackage{graphicx}
\usepackage{textcomp}
\usepackage{xcolor}
\def\BibTeX{{\rm B\kern-.05em{\sc i\kern-.025em b}\kern-.08em
    T\kern-.1667em\lower.7ex\hbox{E}\kern-.125emX}}

\usepackage[letterpaper, top=0.77in, bottom=1.05in, left=0.63in, right=0.63in]{geometry}

\usepackage{algorithm}
\usepackage[colorlinks=true, linkcolor=blue, citecolor=blue, urlcolor=blue]{hyperref}

\def\mrm{\mathrm}
\def\mbf{\mathbf}
\def\beq{\begin{equation}}
\def \eeq{\end{equation}}
\def\bbmat{\begin{bmatrix}}
\def\ebmat{\end{bmatrix}}

\def\pTX{\mbf{p}_{\mrm{TX}}}          % illuminator position
\def\pRX{\mbf{p}_{\mrm{RX}}}          % receiver position
\def\pD{\mbf{p}_{\mrm{D}}}            % target (drone) position
\def\uTX{\hat{\mbf{u}}_{\mrm{TX}}}    % unit vector, target -> illuminator
\def\uRX{\hat{\mbf{u}}_{\mrm{RX}}}    % unit vector, target -> receiver
\def\dvec{\mbf{d}}                    % bisector, \uTX + \uRX
\def\vD{\mbf{v}}                      % target velocity
\def\taub{\tau_{\mrm{b}}}             % bistatic delay
\def\dtau{\Delta\tau}                 % excess delay vs. the direct path
\def\fD{f_{\mrm{D}}}                  % bistatic Doppler shift

\def\Hisac{H_{\mrm{ISAC}}}
\def\Htgt{H_{\mrm{tgt}}}
\def\Hbg{H_{\mrm{bg}}}
\def\sigM{\sigma_{\mrm{M}}}           % mean monostatic RCS, Table 7.9.2.1-1
\def\Nsc{N_{\mrm{sc}}}                % subcarriers on the observation grid
\def\scs{\Delta f}                    % subcarrier spacing
\def\DS{\tau_{\mrm{rms}}}             % RMS delay spread

\begin{document}

% Activate the IEEEtran BIBTEX control configuration
\bstctlcite{IEEEexample:BSTcontrol}

\title{TiamiTwin: A Digital Twin for Bistatic ISAC Drone Sensing, Validated Against Measurements
\thanks{This work was supported by the National Science Foundation (NSF) under Grant CCF-2322191.}
}

%%%% Authors — native IEEEtran form (IEEEauthorblock), the class's own layout.
\author{
\IEEEauthorblockN{Mehdi Zafari\IEEEauthorrefmark{1}\IEEEauthorrefmark{2},
                  Saeede Enayati\IEEEauthorrefmark{2},
                  A. Lee Swindlehurst\IEEEauthorrefmark{1},
                  Amitav Mukherjee\IEEEauthorrefmark{2}}
\IEEEauthorblockA{\IEEEauthorrefmark{1}Department of Electrical Engineering and Computer Science,
                  University of California, Irvine, CA, USA}
\IEEEauthorblockA{\IEEEauthorrefmark{2}Tiami Labs, Sacramento, CA, USA}
}

%%%% Authors — previous authblk version, kept for comparison. To go back to it:
%%%%   1. uncomment \usepackage{authblk} in the preamble,
%%%%   2. comment out the \author{...} block above,
%%%%   3. uncomment the six lines below.
% \author[1,2]{Mehdi Zafari}
% \author[2]{Saeede Enayati}
% \author[1]{A. Lee Swindlehurst}
% \author[2]{Amitav Mukherjee}
% \affil[1]{Department of Electrical Engineering and Computer Science, University of California, Irvine}
% \affil[2]{Tiami Networks}

\maketitle

%%%% Abstract

\begin{abstract}
Monitoring lower airspace over critical infrastructure using cellular signals of opportunity is highly practical because transmitters are pre-deployed, licensed, and continuously active. 
Digital twins can evaluate the feasibility of such integrated sensing and communication (ISAC) architectures, but their predictive accuracy must be validated against real-world data.
% This paper reports the results of such a verification for an operational 5G deployment, with a commercial band n41 gNB and a deployed receiver separated by $572.8\,$m over a non-line-of-sight channel, using measurements  taken from an airborne drone.
This paper reports validation results for TiamiTwin, a digital twin developed for bistatic ISAC drone sensing, using empirical measurements from an operational 5G deployment featuring a commercial band n41 gNB and a receiver separated by 572.8 m over a non-line-of-sight (NLOS) channel. 
% The twin carries three representations of the same channel: the 3GPP TR 38.901 Release 19 bistatic ISAC model, a ray-traced scene of the site, and the measurements themselves, all evaluated on the receiver's $240$-subcarrier grid.
TiamiTwin incorporates three parallel channel representations evaluated on a 240-subcarrier grid: the 3GPP TR 38.901 (Release 19) bistatic ISAC model, a ray-traced site model, and the captured field measurements.
% Both models under-predict the measured root-mean square  delay spread by approximately a factor of three.
Empirical results demonstrate that both statistical and ray-tracing models under-predict the measured root-mean-square (RMS) delay spread by approximately a factor of three. 
% Detection of the drone proves to be a Doppler rather than a power problem, since the drone return lies $68\,$dB below the static zero-Doppler clutter.
% Furthermore, drone target detection is limited by clutter Doppler rather than received power, with target reflections lying 68 dB below static zero-Doppler clutter.
Furthermore, target reflections sit 68~dB below static clutter in power, making target detection entirely dependent on Doppler separation to isolate the drone from zero-Doppler background returns.
% However, the drone location is separable from the clutter along $88\,\%$ of the flight path in delay, Doppler, or both.
Despite this severe clutter environment, the target remains separable along 88\% of the flight path in the delay, Doppler, or joint delay-Doppler domains.
\end{abstract}

\begin{IEEEkeywords}
Integrated sensing and communication (ISAC), digital twins, ray tracing, Sionna, bistatic drone detection, autonomous aerial vehicles, critical infrastructure protection.
% , 5G mobile communication.
\end{IEEEkeywords}

%-------------------------
% Section: Introduction
%-------------------------

\section{Introduction}
\label{sec:intro}

Detecting unauthorized small unmanned aerial systems around critical infrastructure, such as airports, power substations, and military sites, presents a severe operational challenge.
Consumer quadrotors exhibit a radar cross section (RCS) well below the detection threshold of conventional air surveillance radars, whereas deploying, siting, and maintaining dedicated counter-drone sensors for continuous coverage is often cost-prohibitive.
Reusing signals from pre-existing cellular infrastructure offers a scalable, low-cost alternative: a base station (BS) near a protected site continuously radiates licensed signals, allowing a strategically placed passive receiver to detect target-scattered reflections.
Experimental studies have demonstrated drone detection using operational 5G downlinks~\cite{maksymiuk2023passive}, particularly by leveraging the synchronization signal block (SSB) as a sensing waveform, which is periodically broadcast regardless of network traffic load~\cite{jopanya2025ssb}.
Integrated sensing and communication (ISAC) formalizes this %signal reuse
into a designed capability~\cite{gonzalezprelcic2024revolution, rang2026clutter}, accelerating standardization efforts.
Specifically, 3GPP Release 19 introduced Section 7.9 to TR 38.901~\cite{3gpp_tr38901}, defining a bistatic ISAC channel as the superposition of a target response and a background channel encompassing all ambient scatterers.
Consequently, operational sensing deployments can now be evaluated using the same channel modeling frameworks and inter-company calibration benchmarks established for communication systems~\cite{3gpp_r1_2509126}.

%----paragraph 2 v1
A digital twin serves as a cost-effective planning tool for ISAC deployments, as receiver placement and waveform design can be evaluated in simulation far more efficiently than through field trials.
Furthermore, differentiable ray tracing has enabled the creation of site-specific twins with high fidelity~\cite{hoydis2023sionna, wang2025twin}.
However, most published ISAC twins have not been validated against physical site measurements. Instead, standardized models are typically verified against 3GPP reference results~\cite{wu2026implementation, zhao2025buptcmcc, luo2024bistatic}, which confirms specification conformance rather than real-world accuracy at a given site.
Similarly, ray tracers are often calibrated by fitting material parameters in isolation~\cite{ruah2024calibrating}.
% What remains lacking in current literature is a direct comparison of both standardized statistical models and ray-traced digital twins against empirical field measurements evaluated on a common observation grid.
What the current literature lacks is a direct comparison of standardized statistical models and ray-traced digital twins against opportunistic narrowband field measurements from a commercial 5G link, evaluated on a common observation grid.

%----paragraph 2 v2
% A digital twin provides a cost-effective planning tool for ISAC deployments, enabling efficient evaluation of receiver placement and waveform design.
% Furthermore, differentiable ray tracing now enables site-specific twins~\cite{hoydis2023sionna, wang2025twin}.
% However, existing ISAC twins are rarely validated against real physical sites.
% Standardized models are typically verified against 3GPP reference results~\cite{wu2026implementation, zhao2025buptcmcc, luo2024bistatic}, establishing specification conformance rather than site-specific accuracy, while ray tracers rely on fitted material parameters~\cite{ruah2024calibrating}.
% A direct comparison of standardized models and ray-traced twins against real field deployments on a common observation grid remains missing.

%----paragraph 3
Thom\"a et al.~\cite{thoma2026cip} investigated passive sensor placement near protected facilities using multistatic range-Doppler estimation, supported by a 3.75~GHz channel-sounding testbed, a public UAV dataset~\cite{beuster2023testbed}, and experimental evaluations~\cite{beuster2024sounding}.
While their approach relies on dedicated channel-sounding hardware and a complex multistatic architecture, our work evaluates passive sensing feasibility using opportunistic downlink signals from an operational commercial BS and a single deployed receiver.
Similarly, existing 3GPP-based ISAC channel simulators, such as %the one presented in
~\cite{wu2026implementation}, focus on validating Release 19 implementations against theoretical 3GPP calibration benchmarks; in contrast, we perform this calibration and subsequently extend the validation to empirical on-site field measurements.
Furthermore, while Jopanya et al.~\cite{jopanya2025ssb} analyze passive UAV detection via SSB using theoretical Cram\'er-Rao lower bounds in a simulated environment, this paper measures and evaluates real SSB returns collected along an operational 5G link.

%----paragraph 4 v1
In this paper, we evaluate the TiamiTwin framework over an operational non-line-of-sight (NLOS) link spanning 572.8~m between a commercial band n41 5G gNB and a deployed receiver, with an airborne drone operating between them.
The primary contributions of this work are as follows:
\begin{itemize}
\item \textbf{Digital Twin Framework:} Development of a digital twin incorporating three parallel channel representations: the 3GPP TR 38.901 Release 19 model, a ray-traced site scene, and captured field data, all evaluated on the receiver's observation grid.
\item \textbf{Empirical Model Validation:} Demonstration that both standardized stochastic and ray-tracing models under-predict the measured root-mean-square (RMS) delay spread by approximately a factor of three.
\item \textbf{Sensing Feasibility Analysis:} Analysis showing that drone detection relies on Doppler separation to overcome strong static clutter, achieving target separability along 88\% of the flight path.
\item \textbf{Practical Modeling Insights:} Practical modeling takeaways demonstrating that ray-traced delay spread fails to converge with sample budget and exhibits high sensitivity to small endpoint position shifts.
\end{itemize}

\section{System Model and Sensing Geometry}
\label{sec:sys-mod}

\subsection{Geometry and Signal Model}
\label{sec:geom}

%----paragraph 1
An illuminator at $\pTX$ transmits a signal that scatters off a target at $\pD$ and is observed by a receiver at $\pRX$.
Defining $R_{\mrm{T}} = \|\pD - \pTX\|$ and $R_{\mrm{R}} = \|\pRX - \pD\|$ as the two propagation legs, with the direct baseline distance $R_{0} = \|\pRX - \pTX\|$, the scattered return arrives with bistatic delay
\begin{equation}
\taub = \frac{R_{\mrm{T}} + R_{\mrm{R}}}{c},
\qquad
\dtau = \taub - \frac{R_{0}}{c},
\label{eq:delay}
\end{equation}
where $\dtau$ represents the excess delay relative to the direct path.
Resolving $\dtau$ is essential for the receiver to isolate target reflections from ambient clutter and static path signals, aligned with 3GPP TR 38.901 Eq.~(7.9.4-2)~\cite{3gpp_tr38901}.

%----paragraph 2
Two unit vectors point from the target back to the endpoints, $\uTX = (\pTX - \pD)/R_{\mrm{T}}$ and $\uRX = (\pRX - \pD)/R_{\mrm{R}}$.
The angle between them is defined as the bistantic angle $\beta$, and their sum
\begin{equation}
\dvec = \uTX + \uRX,
\qquad
\|\dvec\| = 2\cos(\beta/2),
\label{eq:bisector}
\end{equation}
bisects this angle.
For a target moving with velocity vector $\vD$, the Doppler shift of the scattered path is the projection of $\vD$ onto $\dvec$, given by
\begin{equation}
\fD = \frac{\vD^{\mathsf{T}} \dvec}{\lambda},
\qquad
\left|\fD\right| \le \frac{2\cos(\beta/2)}{\lambda}\,\|\vD\|,
\label{eq:doppler}
\end{equation}
where $\lambda$ is the wavelength.
The factor $\|\dvec\|/\lambda$ defines the Doppler sensitivity in Hz per m/s, attaining its maximum $2/\lambda$ in the monostatic limit ($\beta \to 0^\circ$) and falling to zero in the forward-scatter direction along the baseline ($\beta \to 180^\circ$).
In forward scatter, a target becomes difficult to separate in both delay ($\dtau \to 0$ via~\eqref{eq:delay}) and Doppler ($\|\dvec\| \to 0$ via~\eqref{eq:bisector}), a coupling quantified along a real flight path in Section~\ref{sec:where}.
Because motion orthogonal to $\dvec$ yields no Doppler shift and the endpoints lie on opposite bearings relative to the airborne target, the bisector is oriented nearly vertically. Consequently, a descending drone induces a clear Doppler signature while a level-flying drone does not, though adding a second receiver at a distinct bearing eliminates this degeneracy.

\subsection{The Composite ISAC Channel}
\label{sec:composite}

%----paragraph 3
Section 7.9.4.3 of 3GPP TR 38.901 decomposes the sensing channel into a target contribution and a background contribution~\cite{3gpp_tr38901}, expressed on subcarrier~$f$ as $\Hisac(f) = \Htgt(f) + \Hbg(f)$.
The target channel models the two-hop path from illuminator to target to receiver, whereas the background channel accounts for the direct baseline link and static clutter.
This decomposition enables applying distinct propagation models to each path: for airborne targets, the target legs use the Urban Macro Aerial Vehicle model from TR 36.777~\cite{3gpp_tr36777}, while the background channel uses the terrestrial Urban Macro model from TR 38.901.
Each propagation leg incorporates its own path loss $\mrm{PL}_i$ and log-normal shadow fading $\mrm{SF}_i$.
% Furthermore, the 3GPP path-loss breakpoint distance, which dictates the transition between attenuation slopes, is evaluated assuming a standard terminal height of 1.5~m regardless of the target altitude.
Furthermore, the 3GPP path-loss model is evaluated assuming a standard terminal height of 1.5~m regardless of the target altitude.

%----paragraph 4
Following the procedure in 3GPP TR 38.901 Fig.~7.9.4-1, clusters and rays are generated for each leg according to Section~7.5, concatenated, and scaled for propagation loss and target size~\cite{3gpp_tr38901}.
Denoting $h_{p}$ as the complex coefficient of concatenated path $p$ at delay $\tau_{p}$, the target response on subcarrier $f$ is given by Eq.~(7.9.4-14) of the standard:
\begin{equation*}
\Htgt(f) =
10^{-\frac{\mrm{PL}_1+\mrm{SF}_1+\mrm{PL}_2+\mrm{SF}_2}{20}}
\cdot
\frac{\sqrt{4\pi\sigM}}{\lambda}
\sum_{p} h_{p}\, e^{-j 2\pi \tau_{p} f},
% \label{eq:target}
\end{equation*}
where the first factor accounts for total joint path loss and shadow fading, and the second for target size.
RCS enters the model in two distinct places: a mean monostatic constant $\sigM = -12.81\text{ dBsm}$ applied once per scattering point for a small drone (Table 7.9.2.1-1), and a per-path log-normal variation with a standard deviation of 3.74~dB embedded in $h_{p}$.
Cross-polarization is incorporated per path according to Table 7.9.2.2-1.
Ray concatenation follows Option 2 of Step 9 via a single random pairing across the NLOS rays of both legs, with paths more than 40~dB below the peak response dropped post-concatenation rather than per leg.

%----paragraph 5 v1
In our implementation, $\Hbg$ is modeled as a single-tap Rician channel.
While background link clusters are generated according to Section~7.5 to yield realized delays, powers, and spreads, they are not synthesized into a full multi-tap impulse response.
Instead, all background scattering evaluations reported in this paper are calculated directly from the generated cluster parameters rather than the single-tap $\Hbg$, a distinction that enables reproducibility of the delay and power metrics.

%----paragraph 5 v2
% In our implementation, $\Hbg$ is modeled as a single-tap Rician channel.
% While background link clusters are generated according to Section~7.5 to yield realized delays, powers, and spreads, they are not assembled into a full multipath impulse response.
% All background scattering evaluations reported below are derived directly from this cluster parameter set rather than from $\Hbg$, a distinction essential for result reproducibility.

%-------------------------
% Section: The TiamiTwin Framework
%-------------------------

\section{The TiamiTwin Framework}
\label{sec:twin}

\subsection{Parallel Channel Representations}
\label{sec:three}

\begin{figure}[t]
\centering
\includegraphics[width=0.65\columnwidth]{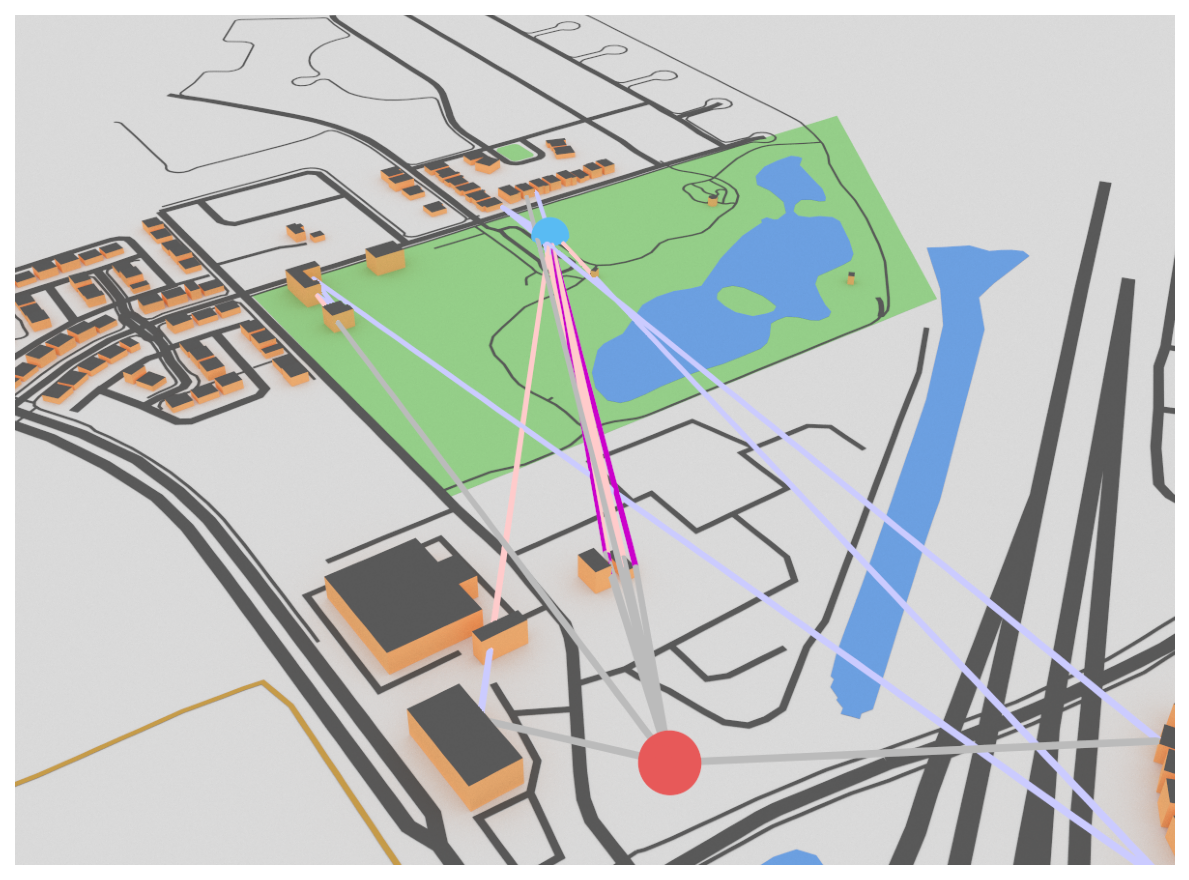}
\caption{Strongest ray-traced paths, from gNB (red) to receiver (blue). No LOS path. The receiver is in the park, which OpenStreetMap renders as bare ground.}
\label{fig:scene}
\end{figure}

%----paragraph 1
TiamiTwin integrates three parallel descriptions of the same physical link.
First, the \emph{stochastic} representation uses the 3GPP model from Section~\ref{sec:composite} to describe an ensemble of urban macro deployments, yielding a distribution essential for determining whether field measurements are typical.
Second, the \emph{ray-traced} representation models the specific site environment using OpenStreetMap building footprints extruded to nominal heights, comprising 436 meshes across approximately 1~$\text{km}^2$.
This model is solved with Sionna RT~\cite{hoydis2023sionna} incorporating specular reflection, diffuse scattering, refraction, and diffraction.
% Object surfaces adopt the ITU-R P.2040~\cite{itu_p2040} parameterization native to Sionna, whereas dielectric parameters for uncovered water and vegetation follow ITU-R P.527-6~\cite{itu_p527}; scattering coefficients are selected as engineering choices because ITU-R P.527-6 specifies permittivity alone.
Static object surfaces are assigned material properties based on ITU-R P.2040~\cite{itu_p2040} and P.527-6~\cite{itu_p527}. Crucially, the target is modeled by a true-to-scale $0.20 \times 0.30$ m 3D drone mesh (DJI Mavic Air equivalent) with ITU-R P.2040 material properties and diffuse scattering enabled.
The drone's small size leads to a sparse set of target-related propagation paths, consistent with the empirical measurements.
% Empirical measurements validate this extreme masking: logged received power (RSRP) shows a negligible $-0.05$ correlation with drone range even when the target passes within $0.5$ m of the receiver, confirming the target is physically buried in the clutter envelope.
Fig.~\ref{fig:scene} illustrates the resulting propagation paths for the studied scenario, highlighting the absence of a line-of-sight (LOS) path.
Finally, the \emph{measured} representation uses empirical data captured along the deployed link, combining channel frequency response estimates extracted from the SSB with drone flight logs.
Integrating these three representations enables TiamiTwin to generate physically grounded channel realizations for unmeasured geometries, justifying the evaluation of its predictive fidelity.

\subsection{Observation Model}
\label{sec:obs}

%----paragraph 2
A meaningful comparison among the three channel representations requires a unified observation grid.
Because empirical estimates are captured over an SSB grid of $\Nsc = 240$ subcarriers with subcarrier spacing $\scs = 30\text{ kHz}$ (a 7.2~MHz bandwidth), the stochastic and ray-traced channels are evaluated on this identical grid rather than over the full carrier bandwidth.
This fixes the delay resolution to $1/(\Nsc \scs) = 139\text{ ns}$.
Although restricting the simulations to the receiver's observation bandwidth sacrifices higher theoretical model resolution, it guarantees that discrepancies between results reflect fundamental channel differences rather than disparate measurement configurations.

%----paragraph 3
Identical processing algorithms are applied across all three channel representations.
Given a channel frequency response $H[k]$ evaluated on subcarriers $k = 0,\dots,\Nsc-1$, the power delay profile (PDP) is
\begin{equation}
p[n] = \Big| \tfrac{1}{\Nsc}\textstyle\sum_{k} H[k]\, e^{\,j 2\pi k n/\Nsc} \Big|^{2},
\qquad
\tau_n = \frac{n}{\Nsc \scs},
\label{eq:pdp}
\end{equation}
and the RMS delay spread is computed as the power-weighted second central moment of this profile:
\begin{equation}
\DS = \sqrt{\frac{\sum_{n \in \mathcal{K}} p[n]\,(\tau_n - \bar{\tau})^{2}}
                 {\sum_{n \in \mathcal{K}} p[n]}},
\
\bar{\tau} = \frac{\sum_{n \in \mathcal{K}} p[n]\,\tau_n}{\sum_{n \in \mathcal{K}} p[n]}.
\label{eq:ds}
\end{equation}
The evaluation index set $\mathcal{K} = \{\, n : p[n] > \max_m p[m] \cdot 10^{-15/10} \,\}$ restricts the calculation to taps within 15~dB of the peak profile power to mitigate noise floor artifacts.

%----paragraph 4 v1
This 15~dB threshold excludes both measurement noise and transform sidelobes inherent to finite-bandwidth profiles, which would otherwise inflate $\DS$ without bound.
Uniformly applying this cutoff enables direct comparisons across all three representations.
Furthermore, because the delay axis in~\eqref{eq:pdp} is periodic, $p[\Nsc-1]$ represents a precursor tap preceding the peak rather than a $33~\mu\text{s}$ delayed response.
Each profile is therefore circular-shifted to align with its peak before computing central moments, yielding consistent measured values between 573 and 675~ns across a 17~dB threshold sweep.
Finally, profiles are averaged over 1-second blocks (50 records) for field measurements, across the statistical ensemble for the 3GPP model, and across eight Monte Carlo solver seeds for the ray-traced scene.

%----paragraph 4 v2
% This 15~dB threshold excludes both measurement noise and transform sidelobes inherent to finite-bandwidth profiles, which would otherwise inflate $\DS$ without bound.
% Applying this cutoff uniformly ensures direct comparability across all three representations. Furthermore, because the delay axis in~\eqref{eq:pdp} is periodic, $p[\Nsc-1]$ represents a precursor tap preceding the peak rather than a $33~\mu\text{s}$ delayed response.
% Each profile is therefore circular-shifted to align with its peak before computing central moments.
% Treating this axis as linear incorrectly registers precursor sidelobes as long-delay echoes, an oversight that artificially inflated 44\% of the 1-second measured blocks into the microsecond regime in uncorrected data.
% With proper circular alignment, the estimator yields consistent measured values between 573 and 675~ns across a 17~dB threshold sweep.
% Finally, profiles are averaged over 1-second blocks (50 records) for field measurements, across the statistical ensemble for the 3GPP model, and across eight Monte Carlo solver seeds for the ray-traced scene.

\subsection{Calibration Against 3GPP Benchmarks}
\label{sec:calib}

%----paragraph 5
To verify implementation fidelity prior to field comparisons, the stochastic model was benchmarked against the 3GPP inter-company calibration campaign for aerial targets~\cite{3gpp_r1_2509126} using the configurations specified in Table 7.9.6.1-1.
Compared to the cross-company average, the median coupling loss differed by 3.6~dB across both frequency bands, while delay spread was 20\% to 23\% higher and azimuth angle-of-arrival spread was 10\% to 20\% lower.
These deviations fall well within the variance reported among contributing 3GPP companies and align with independent evaluations~\cite{wu2026implementation, zhao2025buptcmcc, luo2024bistatic}.
Moreover, comparing the stochastic and ray-traced representations at this geometry yielded target channel coupling loss agreement within 0.16~dB when setting the scene's diffuse scattering coefficient to 0.80.
This tuning represents a single-point alignment rather than a generalizable site property, and was therefore not retained for the field evaluations described next.

%-------------------------
% Section: Results
%-------------------------

\section{Results}
\label{sec:results}

\subsection{Experimental Deployment}
\label{sec:setup}

%----paragraph 1
Field measurements were collected from a commercial 5G site at Tanzanite Community Park in Sacramento, California.
The illuminator is an operational band n41 gNB ($2.50695\text{ GHz}$) with $30\text{ kHz}$ subcarrier spacing, communicating with a receiver comprising two dipole sensors placed $572.8\text{ m}$ away.
The link operates under NLOS conditions, where the earliest measured arrival at $1911\text{ ns}$ closely matches the geometric direct-path delay of $1910\text{ ns}$, corresponding to a diffracted path skirting a building edge.
Surrounding rooftops reach heights up to $21\text{ m}$ against a base station antenna height of $10.5\text{ m}$, characterizing an over-rooftop UMa scenario evaluated from ground level.

%----paragraph 2
The quadrotor target was flown for approximately 22 minutes while the receiver logged SSB channel frequency response estimates at 50 records per second, yielding 65,515 valid records (out of 65,516 captured) after filtering.
The drone operated at 2D ground ranges between 0.5 and 196~m from the receiver, altitudes from 0 to 35~m above ground level, and speeds up to 10.2~m/s.
Two parameters were assumed based on prior site measurements rather than directly logged: a gNB transmit power of 46~dBm and an allocation of 900 resource elements.
Only the absolute signal power level evaluations in Section~\ref{sec:reproduce} depend on these two assumptions.
Moreover, since the receiver operates opportunistically, synchronization relies entirely on the SSB broadcast. Symbol timing and residual carrier frequency offset are estimated and compensated for during PBCH decoding prior to channel estimation.

\subsection{Channel Model Fidelity}
\label{sec:reproduce}

%----paragraph 3
The measured link exhibits high stability: mean received power is $-98\text{ dBm}$ with a standard deviation of $0.6\text{ dB}$, showing negligible correlation ($-0.05$) with drone ground range even during passes within $0.5\text{ m}$ of the receiver.
Because target reflections are not visible in total received power alone, solving the ray-traced scene statically once (rather than dynamically per drone frame) is computationally justified, while further motivating the sensing analysis in Section~\ref{sec:detect}.
The digital twin optimistically predicts a received power of $-86.0\text{ dBm}$, overestimating measured levels by approximately $12\text{ dB}$ (with 10th-to-90th percentile gaps ranging from $+10.5$ to $+13.0\text{ dB}$ across eight solver seeds).
This discrepancy is primarily attributed to unmodeled foliage, as OpenStreetMap supplies building footprints but omits tree canopy.

\begin{figure}[t]
\centering
\includegraphics[width=0.68\columnwidth]{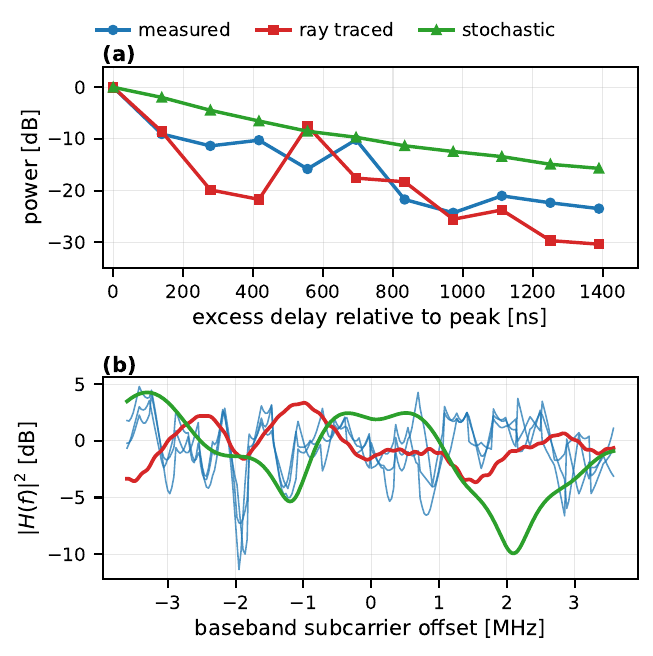}
\caption{Channel comparisons on the receiver $240$ SSB subcarrier
grid: \textbf{(a)} power delay profile, peak aligned; \textbf{(b)} frequency response normalized to its band mean. Both panels share the legend in (a).}
\label{fig:ds}
\end{figure}

%----paragraph 4
Fig.~\ref{fig:ds}(b) demonstrates that all three channel representations exhibit frequency-selective fading of comparable depth ($5$ to $12\text{ dB}$ across the band), but the measured response carries a finer spectral structure corresponding to a larger delay spread.
As shown in Fig.~\ref{fig:ds}(a), the measured PDP decays slowly with rich multipath, whereas the stochastic ensemble is smooth and the ray-traced profile is sparse, as expected for a small set of discrete specular arrivals.
The measured RMS delay spread averages $605\text{ ns}$, tightly distributed with a 10th-to-90th percentile range of $579$ to $640\text{ ns}$ across 1,310 1-second blocks.
In contrast, the stochastic model yields a median RMS delay spread of $232\text{ ns}$ over 2,000 realizations, while the ray-traced scene yields $191\text{ ns}$ averaged across solver seeds, under-predicting empirical measurements by $62\%$ and $68\%$, respectively.
Thus, the physical channel is roughly $2.9\times$ richer in delay spread than predicted by either model.

%----paragraph 5
Although both models similarly under-predict the empirical median, their underlying distributions fundamentally differ.
The stochastic model employs log-normal variability where the measured median delay spread aligns with the 86th percentile, meaning it generates a channel as rich as the physical site roughly 14\% of the time.
In contrast, the ray-traced model never produces a delay spread matching empirical levels.
Additionally, reporting ray-traced results across an ensemble reveals a key methodological finding: ray-traced delay spread fails to converge with sample budget.
Across solver seeds, the 10th-to-90th percentile range spans 93 to 244~ns, reaching a peak of 278~ns at a $10^{9}$ sample budget.
Because the ray-traced profile consists of only a few strong discrete paths, the power-weighted central moment is highly sensitive to diffuse paths fluctuating near the 15~dB threshold.
Nevertheless, because even the maximum ray-traced delay spread (278~ns) remains far below the empirical 605~ns, both models robustly under-predict delay spread regardless of sample count.
In contrast, the dominant path power converges reliably to $-106.6\text{ dB}$ for budgets exceeding $3\times 10^{7}$ rays, confirming that non-convergence is unique to diffuse scattering near the threshold.

%----paragraph 6
Two primary factors account for this under-prediction.
First, the ray-traced scene in Fig.~\ref{fig:scene} incorporates building footprints but lacks foliage, street furniture, and facade details, producing fewer scattering paths and shorter delays (also accounting for the optimistic received power estimate).
Second, the stochastic model relies on generic UMa cluster statistics rather than site-specific geometry.
Neither delay resolution nor estimator sensitivity explains the discrepancy: the 139~ns delay resolution is far finer than all evaluated values (and limited resolution would smooth rather than roughen the profile), while the estimator yields stable measured values between 573 and 675~ns across a 17~dB threshold sweep.
Finally, this comparison specifically evaluates the terrestrial background link, so the bounds on aerial model validity noted in Section~\ref{sec:where} do not apply.

\subsection{Target Sensing Feasibility}
\label{sec:detect}

%----paragraph 7
Target detection in this scenario hinges on Doppler separation rather than signal power.
To show this, we evaluate a simulation with the drone positioned 100~m above ground level (above the rooftop canopy), unlike the lower-altitude experimental flight analyzed in Section~\ref{sec:where}.
Target detection via power alone is infeasible because only three of the 192,000 simulated ray paths interact with the target, yielding a combined target reflection 68.4~dB below the static background clutter.
Detection must therefore rely on Doppler processing.
While static clutter paths exhibit zero Doppler shift, the target geometry yields a bistatic angle $\beta = 143.5^{\circ}$ in~\eqref{eq:doppler}, producing $\|\dvec\| = 0.63$ with the bistatic bisector directed $89^{\circ}$ below horizontal.
Thus, $\dvec$ is almost purely vertical, with its horizontal projection having a magnitude of only $0.0099$ ($63\times$ smaller).

\begin{figure}[t]
% \vspace{5pt}
\centering
\includegraphics[width=0.68\columnwidth]{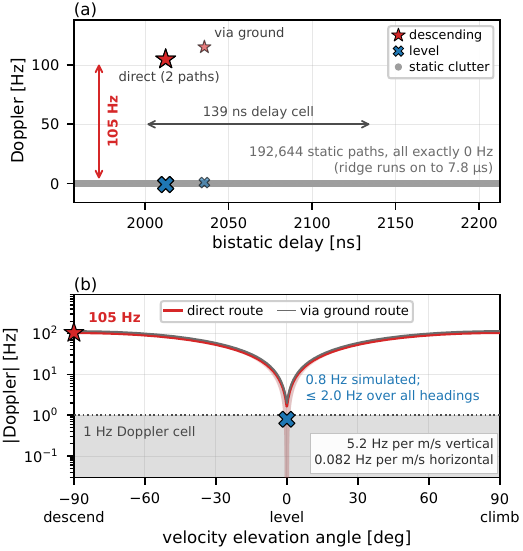}
\caption{\textbf{(a)} Delay and Doppler plane for a drone descending and flying level at
$20\,$m/s; every static path sits at exactly zero Doppler. \textbf{(b)} $|f_D|$ versus
velocity elevation, the band covering all headings, against the $1\,$Hz cell.}
\label{fig:dd}
\end{figure}

%----paragraph 8
% Fig.~\ref{fig:dd}(a) directly contrasts these motion profiles: a target descending at $20\text{ m/s}$ induces a $+104.8\text{ Hz}$ Doppler shift, separating cleanly from zero-Doppler clutter.
Fig.~\ref{fig:dd}(a) contrasts these motion profiles. A target descending at $20\text{ m/s}$ physically induces a $+104.8\text{ Hz}$ Doppler shift. Because the 50 record/s SSB capture rate imposes a $\pm 25$ Hz unambiguous window, this shift aliases to $+4.8\text{ Hz}$, well outside the $1\text{ Hz}$ zero-Doppler clutter cell.
Conversely, level flight at the same speed yields at most $2.0\text{ Hz}$ across all propagation routes and headings (and only $-0.8\text{ Hz}$ for the heading evaluated in ray tracing), with ray-traced and analytical calculations agreeing within $0.05\text{ Hz}$.
Fig.~\ref{fig:dd}(b) generalizes this behavior across all 3D velocity directions in~\eqref{eq:doppler}.
Doppler sensitivity is $5.24\text{ Hz per m/s}$ for vertical motion versus only $0.083\text{ Hz per m/s}$ for horizontal motion.
Therefore, clearing a $1\text{ Hz}$ Doppler resolution cell requires $0.19\text{ m/s}$ of vertical rate, compared to $12.1\text{ m/s}$ of level flight along the most favorable heading; velocity elevation angles exceeding $1.5^{\circ}$ guarantee cell transition regardless of heading.
Thus, level flight is largely masked within the zero-Doppler clutter region, whereas vertical motion is unambiguously detectable.
Finally, note that $\|\dvec\| = 0.63$ yields only $31\%$ of the equivalent monostatic Doppler shift ($2\|\vD\|/\lambda$), even for vertical trajectories.

%----paragraph 9 v1
Finally, target visibility depends deterministically on local geometry rather than  solver randomness.
For instance, shifting the receiver by 4.5~m drops the target-to-clutter ratio by 12~dB (from $-56.5$ to $-68.4\text{ dB}$) as specular facade paths disappear, whereas solver reseeding produces negligible variation.
Ray-traced target returns must therefore be evaluated against specific geometric boundaries rather than aggregate path counts.

%----paragraph 9 v2
% Another key methodological insight is that target visibility exhibits extreme sensitivity to exact spatial geometry.
% Displacing the receiver by merely 4.5~m (holding frequency, solver seed, and ray budget constant) reduces the target path count from nine to three, dropping the target-to-clutter power ratio from $-56.5\text{ dB}$ to $-68.4\text{ dB}$.
% Conversely, reseeding the solver for a fixed geometry alters the path count by at most one path.
% This variation is purely geometric rather than stochastic, arising when specular reflections off finite building facades fail to satisfy ray-path conditions as endpoints shift.
% Consequently, reporting target visibility relative to clutter requires explicitly citing the underlying geometry and propagation mechanisms rather than relying on aggregate path counts.

\subsection{Spatial Target Separability}
\label{sec:where}

\begin{figure*}[t]
\centering
\includegraphics[width=0.7\textwidth]{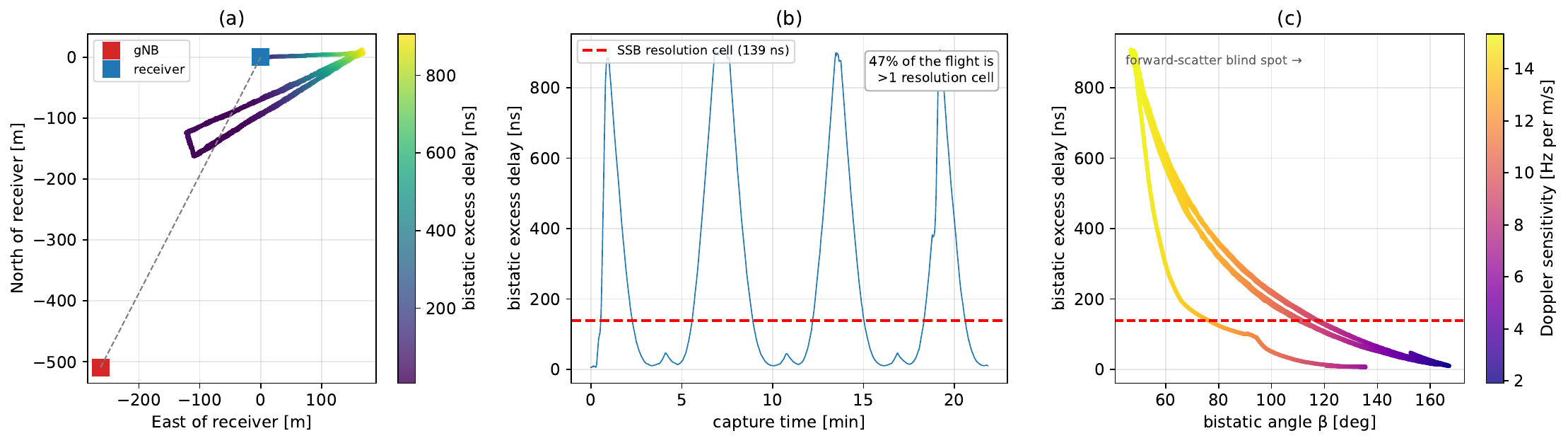}
\caption{Bistatic geometry at every measured drone position: \textbf{(a)} the track,
\textbf{(b)} excess delay against time, \textbf{(c)} excess delay against bistatic angle, colored by Doppler sensitivity. In (c) both separability axes fall together as
$\beta \to 180^{\circ}$, sharing that angle.}
\label{fig:track}
\end{figure*}

%----paragraph 10
% Evaluating the experimental drone flight, with tracks and geometry visualized in Fig.~\ref{fig:track}, against the geometry in Section~\ref{sec:geom} reveals significant spatial variation in target detectability.
Applying the geometric framework from Section~\ref{sec:geom} to the experimental flight track visualized in Fig.~\ref{fig:track} reveals significant spatial variation in target detectability.
The target's bistatic excess delay spans $6$ to $907\text{ ns}$, allowing $47\%$ of the flight path to be separated from static clutter based solely on the receiver's $139\text{ ns}$ delay resolution.
Doppler sensitivity along the track ranges from $1.91$ to $15.34\text{ Hz per m/s}$ (median $8.24\text{ Hz per m/s}$), producing Doppler magnitudes up to $152\text{ Hz}$ (median $11.5\text{ Hz}$) at measured speeds.
To establish joint performance bounds, each flight record is evaluated against a $139\text{ ns}$ delay cell and a $1\text{ Hz}$ Doppler cell (corresponding to $1\text{ s}$ coherent integration, matching Section~\ref{sec:reproduce}).
Target velocities are derived by differentiating logged positions, with Doppler shifts folded into the $\pm 25\text{ Hz}$ unambiguous window imposed by the $50\text{ record/s}$ rate; aliased shifts landing on zero-Doppler clutter are classified as hidden.

\begin{figure*}[t]
\centering
\includegraphics[width=0.7\textwidth]{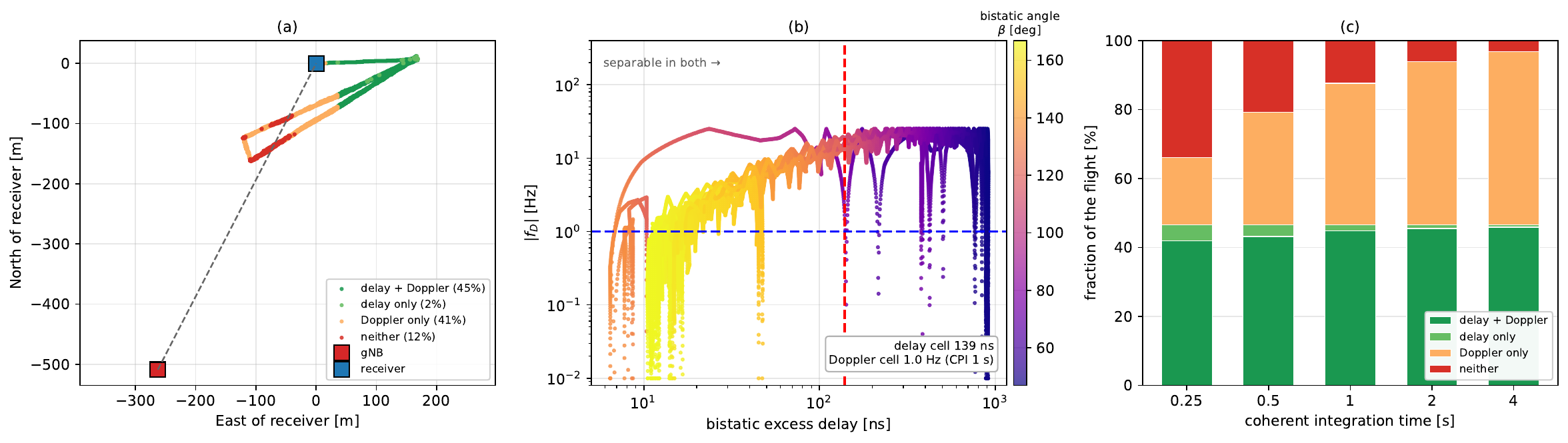}
\caption{Each record classified against both resolution cells, $139\,$ns and
$1\,$Hz: \textbf{(a)} on the track, \textbf{(b)} in the joint plane, \textbf{(c)} against integration time. Doppler separability is more available than delay, leaving $12\,\%$ dark in both. Shifts are folded into the $\pm25\,$Hz window, so aliasing counts as hidden.}
\label{fig:joint}
\end{figure*}

%----paragraph 11
As shown in Fig.~\ref{fig:joint}, joint classification reveals that $45\%$ of the flight path is separable in both delay and Doppler domains, $41\%$ in Doppler alone, $2\%$ in delay alone, and $12\%$ in neither, yielding an overall target separability of $88\%$.
While forward-scatter geometries create extensive delay blind spots, Doppler processing successfully recovers $53\%$ of the trajectory unreachable by delay alone.
Furthermore, although $21.5\%$ of records experience Doppler aliasing beyond the unambiguous window, folded shifts land clear of the zero-Doppler clutter ridge and remain detectable.
The remaining $12\%$ ``dark zone" occurs when the target hovers below $0.5\text{ m/s}$ or flies level at forward-scatter bistatic angles ($\beta \in [130^{\circ}, 167^{\circ}]$), where excess delay and Doppler sensitivity collapse simultaneously per~\eqref{eq:delay} and~\eqref{eq:bisector}.
Ray-tracing evaluations along actual track positions confirm this behavior, resolving 3 to 4 target returns for $\beta \in [48^{\circ}, 56^{\circ}]$ but zero returns at $\beta = 164^{\circ}$.
It is critical to note this analysis represents a geometric upper bound on domain separability, not a realized detection rate.
Because the target sits roughly 68 dB below the clutter floor, actual detection requires sufficient link budget and advanced clutter suppression, which a baseline energy detector cannot overcome.

%----paragraph 11
Varying coherent integration time ($T_{\text{CPI}}$) highlights fundamental domain trade-offs: increasing $T_{\text{CPI}}$ from $0.25$ to $4\text{ s}$ expands Doppler-only separability from $66\%$ to $97\%$, but joint delay-Doppler separability increases only modestly from $42\%$ to $46\%$ because delay resolution remains bandwidth-bound.
For site planning, a single bistatic pair inherently exhibits a predictable dark zone along its baseline covering roughly $10\%$ of a trajectory; this blind spot can be compressed by longer dwell times or eliminated entirely by deploying a second receiver on a distinct bearing.
Finally, 3GPP TR 36.777 path loss models are specified for target altitudes above $22.5\text{ m}$, whereas the evaluated flight operates mostly below this threshold where the model extrapolates.
While all position-based geometric metrics (delay, bistatic angle, Doppler sensitivity) remain exact, stochastic target path loss predictions at low altitudes should be interpreted as optimistic.

%-------------------------
% Section: Conclusion
%-------------------------

\section{Conclusion}
\label{sec:conc}

TiamiTwin evaluates an operational bistatic ISAC link by unifying a 3GPP Release 19 stochastic model, a ray-traced digital twin, and empirical measurements on a common grid.
While accurately capturing sensing geometry, both theoretical models systematically under-predict empirical delay spread by nearly a factor of three and overestimate received power by $12\text{ dB}$.
Subsequent multi-site validations confirm that these limitations generalize broadly; standard models predict identical delay spreads for both heavily cluttered parks and barren line-of-sight environments.
Because target reflections sit $68\text{ dB}$ below static clutter, practical target separation relies strictly on Doppler processing.
Consequently, synthesizing detectors solely from theoretical models risks assuming unrealistically sparse channels.
This motivates data-driven channel learning and multistatic architectures to eliminate geometric blind spots as critical future work.

% References
%-----> \balance evens out the two columns on the last page. Enable it once the
%       paper is at its final length; running it on an under-full last page can
%       error ("balance: no floats allowed").
% \balance
\bibliographystyle{IEEEtran}
\bibliography{ref}

\end{document}